\UseRawInputEncoding 
\documentclass[a4paper]{jpconf}
\usepackage{graphicx}
\usepackage{amsfonts}
\usepackage{bbm}
\usepackage{epsfig}
\usepackage{times}
\usepackage[english]{babel}
\usepackage{color}
\usepackage{hyperref}
\usepackage{framed}
\usepackage{changes}
\usepackage{physics}
\usepackage{mathtools}
\usepackage{amsthm}
\usepackage{amsmath}
\usepackage{amssymb}
\usepackage{tensor}
\usepackage{float}
\usepackage{color}
\usepackage[caption=false]{subfig}

\newcommand{\eq}{\begin{equation}}
\newcommand{\qe}{\end{equation}}
\newcommand{\lp}{\left(}
\newcommand{\rp}{\right)}
\newcommand{\mbf}[1]{\mathrm{#1}}

\begin{document}
\title{Fall-to-the-centre as a $\mathcal{PT}$ symmetry breaking transition}

\author{Sriram Sundaram}

\address{Department of Physics and Astronomy, McMaster University, 1280 Main St. W., Hamilton, Ontario, Canada, L8S 4M1\\}

\author{C. P. Burgess}
\address{Department of Physics and Astronomy, McMaster University, 1280 Main St. W., Hamilton, Ontario, Canada, L8S 4M1\\}
\address{Perimeter Institute for Theoretical Physics, 31 Caroline St. N., Waterloo, Ontario, Canada, N2L 2Y5\\}

\author{D. H. J. O'Dell}
\address{Department of Physics and Astronomy, McMaster University, 1280 Main St. W., Hamilton, Ontario, Canada, L8S 4M1\\}

\ead{dodell@mcmaster.ca}

\begin{abstract}
The attractive inverse square potential arises in a number of physical problems such as a dipole interacting with a charged wire, the Efimov effect, the Calgero-Sutherland model, near-horizon black hole physics and the optics of Maxwell fisheye lenses. Proper formulation of the inverse-square problem requires specification of a boundary condition (regulator) at the origin representing short-range physics not included in the inverse square potential and this generically breaks the Hamiltonian's continuous scale invariance in an elementary example of a quantum anomaly. The system's spectrum qualitatively changes at a critical value of the inverse-square coupling, and we here point out that the transition at this critical potential strength can be regarded as an example of a $\mathcal{PT}$ symmetry breaking transition. In particular, we use point particle effective field theory (PPEFT), as developed by Burgess \textit{et al} \cite{cliffppeft1}, to characterize the renormalization group (RG) evolution of the boundary coupling under rescalings. While many studies choose boundary conditions to ensure the system is unitary, these RG methods allow us to systematically handle the richer case of nonunitary physics describing a source or sink at the origin (such as is appropriate for the charged wire or black hole applications). From this point of view the RG flow changes character at the critical inverse-square coupling, transitioning from a sub-critical regime with evolution between two real, unitary fixed points ($\mathcal{PT}$ symmetric phase) to a super-critical regime with imaginary, dissipative fixed points ($\mathcal{PT}$ symmetry broken phase) that represent perfect-sink and perfect-source boundary conditions, around which the flow executes limit-cycle evolution. 
\end{abstract}

\section{Introduction}
In the presence of an attractive $1/r$ potential a classical particle will follow either an elliptic, hyperbolic or parabolic trajectory. However, in more singular potentials the particle can exhibit a new type of behaviour where it spirals down onto the origin, a phenomenon called  ``fall to the centre'' \cite{lifshitz1981quantum,case1950singular,perelomov1970faltothecenter}.  The least singular potential where this occurs is the attractive inverse square potential $-g/r^2$ which has precisely the same radial dependence as the centrifugal barrier and hence can overcome it for large enough $g$.

The behaviour of quantum particles in an attractive inverse square potential has been studied in an experiment by Denschlag, Umshaus and Schmiedmayer \cite{schmiedmayer1998coldatom} who scattered cold lithium atoms from a thin (radius $\sim 1 \mu$m) charged wire and observed fall to the centre. The existence of the inverse square potential in this case is easily understood:
consider a neutral but polarizable atom interacting with a charged wire; the radial electric field $\mathcal{E}$ emanating from the wire falls off as $1/r$ and induces a dipole moment $d$ in the particle that in the linear response regime is proportional to the strength of the field. The interaction energy $- (1/2) d \mathcal{E}$ between the atom and the wire must then go as the inverse square $-1/r^2$ of the distance between the atom and the wire. The Schr\"{o}dinger equation describing this situation takes the form
\begin{equation}
 -\frac{\hbar^{2}}{2m}\frac{\partial^{2}\psi}{\partial Q^2} - \frac{g}{Q^{2}}\psi = \mathrm{i}\hbar\frac{\partial\psi}{\partial t} 
 \label{eq:timedepSchrod}
\end{equation}
(since in this paper we will only consider one-dimensional problems we have replaced the radial coordinate $r$ with the coordinate $Q$ which lies in the range $-\infty \le Q \le \infty$). Other physical situations where the inverse square potential appears include the Efimov effect (a counter-intuitive family of bound states of three particles with an infinite bound state spectrum given by a geometric series) \cite{efimov1970,efimov1973,braaten2006,Kraemer2006,Pollack2009,Ferlaino2010,bhaduri2011efimov,Huang2014,Pires2014,Tung2014,moroz2015efimov,kunitski2015}, Calogero-Sutherland models \cite{pkp1999equivalence}, and near the event horizon of black holes \cite{srinivasan1999particle,Gupta2001horizon,camblong2003}. Analogue inverse square potentials also appear in various ways in optics, such as in the optical coherence of sunlight \cite{sundaram2016origin}, and in Maxwell fisheye lenses where the refractive index  takes an inverse square form  \cite{alonso15,hidden_symmetry_fisheye_frank1990,leonhardt2009perfect}.

The peculiar properties of inverse square potentials in quantum mechanics have been widely discussed in the literature, and good introductions can be found in Refs.\ \cite{holstein2002anomalies,griffiths2006}. The key feature of Eq.\ (\ref{eq:timedepSchrod}) is that it is scale invariant under joint continuous scaling of space and time, $Q \rightarrow s Q$ and $t \rightarrow s^{2}t$. This means that, unlike the Coulomb potential where the Bohr radius provides a length scale, there is no  natural length scale in the inverse square problem. Indeed, even if we consider the one dimensional case (as we do here) such that there is no centrifugal barrier, the system is still saved from collapse in the Coulomb case by the zero-point kinetic energy $\sim \frac{\hbar^2}{2m a^2}$ associated with a state of size  $a$ beating the potential energy $-\frac{q_{1}q_{2}}{4 \pi \epsilon_{0} a}$ at small enough $a$. The same is \textit{not} true in the inverse square case where both the zero-point energy and the potential scale in the same way. In fact, if we can find one solution of Eq.\ (\ref{eq:timedepSchrod})  then we have found an infinite family of them related by $s$. Thus, if we can find one bound state with (negative) energy $E$ then there is a continuum of bound states $s^{2} E$ with every possible negative energy and the spectrum is therefore unbounded from below.

\begin{figure}
\begin{center}
\includegraphics[width = 0.5\columnwidth]{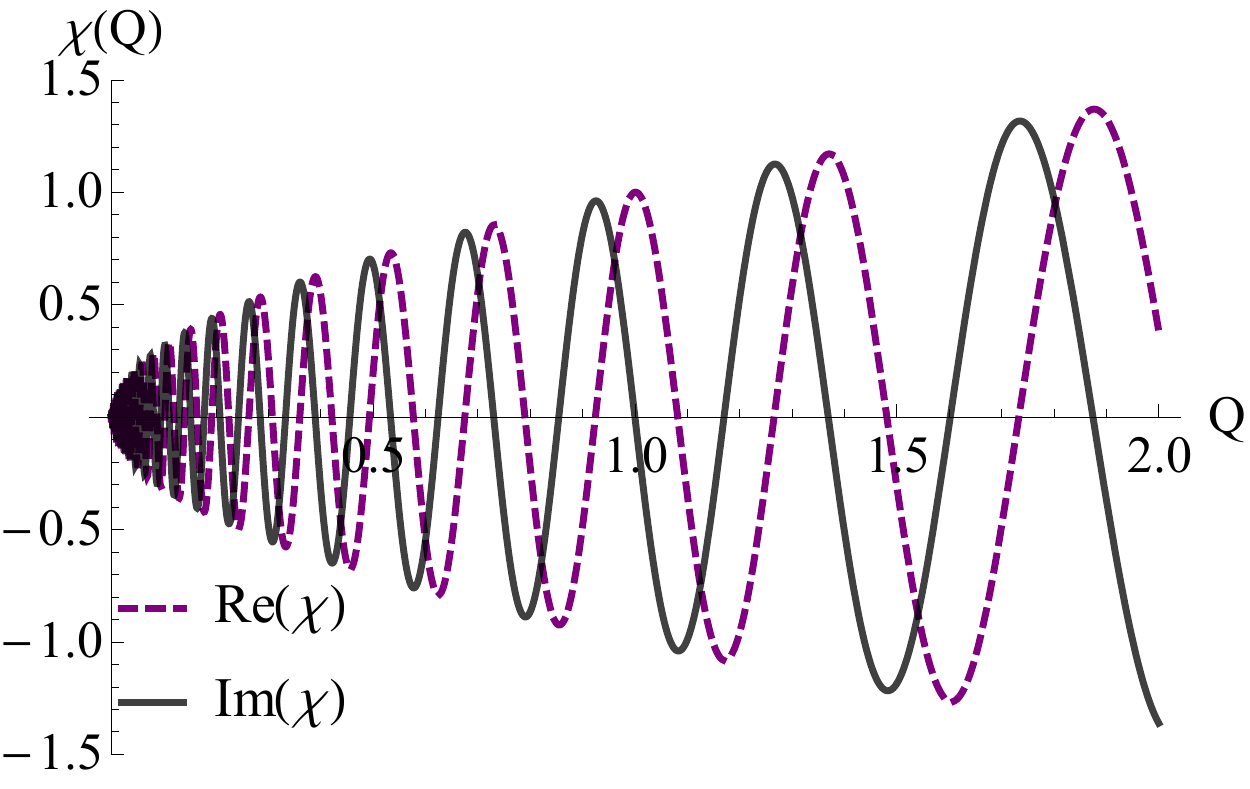}
\end{center}
\caption{The wavefunction $Q^{-i \vert \sigma \vert + 1/2}=Q^{1/2}e^{-i \vert \sigma \vert \log Q}$ in the supercritical regime with $\vert \sigma \vert = 20$. Note the logarithmic phase singularity at Q = 0 where the phase is undefined because it oscillates so fast it takes all values at once. The purple dashed line shows the real part  and the black solid line shows the imaginary part. }
\label{fig1}
\end{figure}

Despite this pathology, Eq.\ (\ref{eq:timedepSchrod}) has exact stationary solutions $\psi(Q,t)=\chi(Q) \exp (-\mathrm{i} E t/\hbar)$, where $\chi(Q)$ are Hankel functions in the case of scattering states ($E>0$) and modified Bessel functions in the case of bound states ($E<0$). The problem lies, however, in finding appropriate boundary conditions, i.e.\  the linear combination of these solutions that describes a particular physical situation. In the Coulomb problem we are able to choose one of the two linearly independent solutions to the radial equation simply by its asymptotic behaviour: at small distances where the centrifugal barrier dominates, the solution to the hydrogenic radial equation with angular momentum quantum number $l$ (an integer) takes the form
\begin{equation}
u(r)=A \ r^{l+1}+B \ r^{-l}
\end{equation}
and since $r^{-l}$ blows up as $r \rightarrow 0$ (assuming $l \neq 0$) we can set $B=0$ without further thought. The same logic cannot in general be applied to the inverse square potential. At small distances where we can ignore the eigenvalue $E$ in comparison to the other terms in the eigenvalue equation, the solutions are of the form
\begin{equation}
 \chi(Q) = C_{+}~Q^{1/2+ \sigma} + C_{-}~Q^{1/2- \sigma} 
 \label{eq:shortrange_chi} 
\end{equation}
where $\sigma = \sqrt{1/4- \alpha}$ and $\alpha \equiv  2m g /\hbar^2$. Defining $\alpha_{c} \equiv 1/4$, which is the critical value at which fall to the centre first takes place, we see that when $\alpha < \alpha_{c}$ (subcritical regime) the two solutions are distinguishable in terms of their behaviour as $Q \rightarrow 0$, but when
$\alpha > \alpha_{c}$ (supercritical regime) the modulus of both solutions is identical and they only differ by a phase (see Figure \ref{fig1}), and there is no simple criterion like boundedness at the origin for choosing one solution over the other. Clearly we need to supply a boundary condition to fix the ratio $C_{+}/C_{-}$; the fact we did not need to do this in the Coulomb case is a reflection of the latter's rather special properties (superintegrability). Furthermore, the logarithmic phase divergence present in both solutions in Eq.\ (\ref{eq:shortrange_chi}) indicates that our problem is missing some physics near the origin.

\begin{figure}
\begin{center}
\includegraphics[width = 0.5\columnwidth]{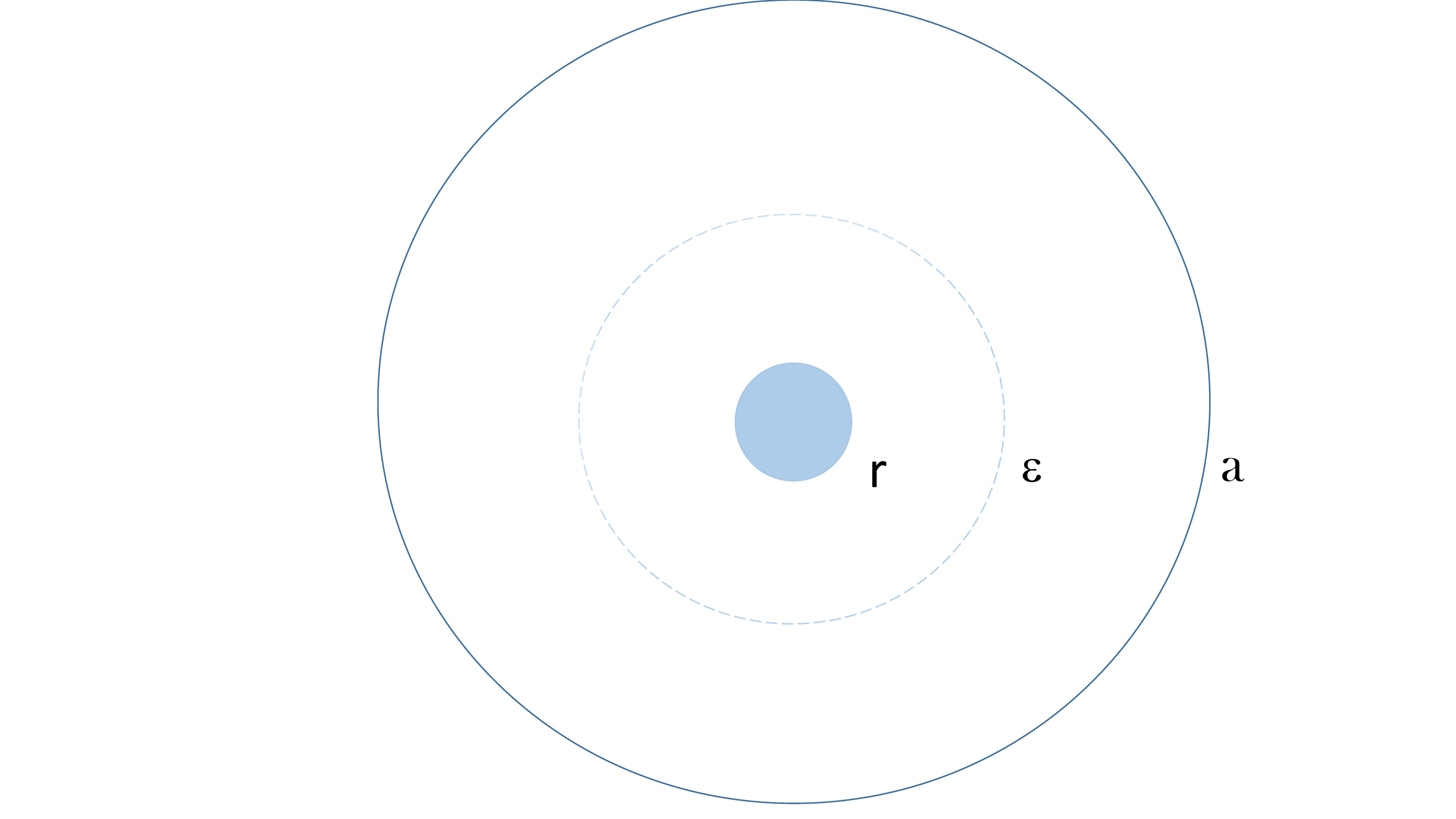}
\end{center}
\caption{A figurative illustration of the hierarchy of length scales in the regularization of the fall-to-the-centre problem. The length $a$ is the scale we have access to in experiments which is very large compared to the regulator scale $\epsilon$ which is in turn very large compared to $r$, the size of the source located at the origin, i.e.\ $r \ll \epsilon \ll a$. The boundary condition is derived using the PPEFT action that describes the properties of the source at the regulator scale $\epsilon$ (which is arbitrary). Renormalizing the source-bulk coupling ensures that physical predictions at scale $a$ are independent of the regulator $\epsilon$. }
\label{fig2}
\end{figure}

A resolution to these issues is suggested by the atom-wire problem. The inverse square potential is merely the long range behaviour of the atom-wire  interaction and when they approach closely the fact that the wire has a finite width becomes important. This introduces the microscopic length scale $r$  (the radius of the wire), see Figure \ref{fig2}, that provides a short-distance cut-off that regulates the singularity in the inverse square potential and breaks the continuous scaling symmetry.  The breaking of classical continuous scale invariance in a quantum theory is termed a \textit{scale anomaly} \cite{holstein2002anomalies}.  Although anomalies were originally conceived in the context of particle physics, they have recently been realized in ultracold atomic systems through the Efimov effect   \cite{Kraemer2006,Pollack2009,Ferlaino2010,bhaduri2011efimov,Huang2014,Pires2014,Tung2014,moroz2015efimov,kunitski2015}, and frequency shifts of breathing modes in trapped 2D Fermi gases \cite{pitaevskii1997breathing,olshanii2010anomaly,taylor2012,holten2018,peppler2018,murthy2019}. A scale anomaly has also been observed in graphene where the relativistic nature of the dispersion relation means that it is the Coulomb potential that leads to a scale invariant system in that case \cite{anomalygraphene2017ovdat}.

The `high-energy' physics at the microscopic length scale $r$ might be complicated, and we might even be ignorant of its detailed form, but if all we can observe is the low-energy behaviour at large distances then we can treat the problem using the techniques of effective field theory which provide systematic methods for regulating the theory. The version that we apply in this paper is known as point particle effective field theory (PPEFT)  and allows us to specify the boundary condition in an intuitive and physical way through the presence of an imagined point `particle' that sits the origin \cite{cliffppeft1,plestid2018fall}.  The theory is conveniently formulated in terms of the action $S_{\mathrm{total}}=S_{B} + S_{p}$, where $S_{B}$ is the action for the `bulk' field $\chi$ that has Eq.\ (\ref{eq:timedepSchrod}) as its equation of motion, and $S_{p}$ is the point particle effective action that describes the microscopic physics localized at the origin and how it couples to the bulk field.
 In this paper we will review the application of PPEFT to the inverse square problem and confirm that at lowest order the effect of the microscopic physics on the inverse square problem is simply to add a {\it compulsory} Dirac $\delta$-function to  Eq.\ (\ref{eq:timedepSchrod}) so that the time-independent Schr\"{o}dinger equation reads 
\begin{equation}
 -\frac{d^{2}\chi}{dQ^2} - \frac{\alpha}{Q^{2}}\chi + \lambda \delta(Q) \chi= k^2 \chi 
 \label{eq:schrod_with_delta}
\end{equation}
where $k^2=2mE/\hbar^2$.
By integrating this equation over an infinitesimal range $-\epsilon < Q < \epsilon$ we can see that the value of the coupling constant $\lambda$ determines the jump in gradient of $\chi$ at the origin and provides the necessary boundary condition at the origin. This boundary condition is unitary (with hermitian Hamiltonian) when $\lambda$ is real, but is generically not unitary if $\lambda$ is complex.

The boundary condition fixes the ratio of integration constants, $C_{+}/C_{-}$, and this ratio determines the problem's intrinsic length scale: the ratio of the solutions in Eq.\ (\ref{eq:shortrange_chi}) gives the length scale $L=(C_{+}/C_{-})^{-\frac{1}{2 \sigma}}$ which breaks the scale invariance as long as both $C_{+}$ and $C_{-}$ are finite. 
In this picture the coupling $\lambda$ is to be regarded as depending on the position $\epsilon$ at which the boundary condition is imposed in such a way as to ensure that the physical scale $C_+/C_-$ remains fixed. This defines a renormalization-group (RG) flow for $\lambda$ under scale transformations. The interpretation of this flow is clearest when $\sigma$ is real, i.e.\ $\alpha<\alpha_{c}$. Through the boundary condition $\lambda$ controls the relative weight of the two solutions $Q^{1/2 \pm \sigma}$ at $Q = \pm \epsilon$. This relative weight changes as $Q$ changes, with the $Q^{1/2+\sigma}$ solution eventually dominating at large length scales, corresponding to an infrared (IR) fixed point for which the wavefunction will become scale invariant. In other words, we flow towards a scale invariant IR fixed point associated with $C_{-}=0$. Conversely, in the opposite ultraviolet (UV) limit we have a scale invariant fixed point for which $Q^{1/2-\sigma}$ dominates, associated with $C_{+}=0$. These simple arguments will be made more concrete in the rest of this paper, but the upshot is that  $\lambda$ depends on the energy/length scale of our observations, i.e.\ undergoes renormalization \cite{gupta1993renormalization,holstein2002anomalies,camblong2003,mueller2004renormalization,kaplan2009conformality,moroz2010nonrelativistic,cliffppeft1}. 

The full renormalization group (RG) flow of $\lambda$ with scale is nonlinear. In the subcritical regime there are two fixed points where the solution is scale invariant, as argued above. The UV fixed point is unstable, the IR one is stable, and the flow between them can in general pass through complex values of $\lambda$ (see Figures \ref{fig3} and \ref{fig4}) so that the Hamiltonian in Eq.\ (\ref{eq:schrod_with_delta}) is generically non-hermitian, although in the special case where the flow starts on the real axis $\lambda$ will always remain real.   Either way, all trajectories in the subcritical regime eventually tend to the real value of $\lambda$ at the IR fixed point. However, as $\alpha$ increases the two fixed points approach one another and merge exactly at $\alpha=\alpha_{c}$; thereafter they proceed to evolve as complex conjugates in the complex plane \cite{kaplan2009conformality}. This heralds a topological change in the RG flow such that the trajectories become limit cycles that can never reach the fixed points, see Figure \ref{fig4}. The evolution of $\lambda$ along a limit cycle is periodic on a logarithmic scale and this implies that in the supercritical regime continuous scale invariance is broken in favour of a discrete version (quantum anomaly). This is the explanation for the geometric series bound state spectrum found in the Efimov  problem \cite{braaten2006}, although the Efimov case is exceptional because $\lambda$ is real and this corresponds to a self-adjoint extension of the original inverse square problem. The general case again corresponds to complex values of $\lambda$ that give a complex extension of the supercritical inverse square problem and have been used to describe inelastic scattering \cite{moroz2010nonrelativistic} and the absorption of particles on a charged wire \cite{plestid2018fall}.  Close examination of the trajectories in Figure \ref{fig4}  shows that the flow near the real $\lambda$ axis in the supercritical regime is unstable because it forms a separatrix between trajectories in the upper and lower half planes such that a small deviation can lead to very different flow (around one fixed point or the other). In fact, the closer the trajectory approaches the real axis during that part of its motion, the further it gets swept away during the rest. Of the two complex stationary points, one corresponds to a perfect absorber of probability, and the other a perfect emitter of probability.



One of the articles of faith of quantum mechanics used to be that the Hamiltonian should be hermitian: this leads to a real energy spectrum and guarantees unitary time evolution where probability is preserved.  However, starting in 1998 it was pointed out by  Bender and coworkers \cite{bender1998real, bender1999pt, bender2002complex} that non-hermitian Hamiltonians possessing $\mathcal{PT}$ symmetry can have exclusively real eigenvalues. $\mathcal{P}$ is the parity operator that effects the transformation $Q \rightarrow -Q$, and $\mathcal{T}$ is the time-reversal operator that effects the transformations $t \rightarrow -t$ and $\mathrm{i} \rightarrow -\mathrm{i}$.  Furthermore, there exists a phase transition as a function of a parameter where some of the eigenstates spontaneously break $\mathcal{PT}$ symmetry and the corresponding eigenvalues become complex.  This observation has led to an explosion of interest in non-hermitian Hamiltonians in quantum mechanics  \cite{bender2003must, bender2010pt, das2009alternative, das2009structure, dorey2001supersymmetry, abhinav2010supersymmetry, abhinav2013conserved,mandal2013pt,mandal2015pt},  and optics \cite{lumer2013nonlinearly,zyablovsky2014pt, raja2019multifaceted,raja2020tailoring,longhi2018parity}, including experimental confirmation  \cite{bender2013observation, schindler2011experimental,guo2009observation,ruter2010observation}.  

In a far-sighted paper on RG fixed point mergers, Kaplan \textit{et al.}  \cite{kaplan2009conformality} point out that the Berezinskii-Kosterlitz-Thouless (BKT) phase transition (vortex-antivortex pair unbinding transition) is also an example of a phase transition where conformal scaling is lost when two real fixed points merge and enter the complex plane, exactly as happens in the inverse square potential problem. In this paper we also focus on the behaviour of the fixed points of the RG flow and suggest that it is analogous to Bender's $\mathcal{PT}$  symmetry breaking transition. Indeed,  the Hamiltonian  in Eq.\ (\ref{eq:schrod_with_delta}) is typically non-hermitian and undergoes a transition as a function of the parameter $\alpha$ from the subcritical regime, where the flow is organized by real fixed points, to the supercritical regime where it is organized by complex ones.  Whereas the real fixed points display $\mathcal{T}$ symmetry, this is broken when they become complex (one representing a source, the other a sink of probability). The eigenfunctions $Q^{1/2 \pm \sigma}$ also change their nature when $\alpha > \alpha_{c}$, going from being real to complex and developing a phase singularity at the origin.  While not all aspects of the connection are clear to us at the time of writing (such as the role of $\mathcal{P}$) the basic scenarios are similar enough to warrant investigation.



The rest of this paper is organized as follows :  in Section \ref{sec:inversesqH} we recapitulate the inverse square Hamiltonian and mention a few points not already covered in the Introduction.  In Section \ref{sec:ppeft} we introduce PPEFT and use it to  derive a boundary condition due to a microscopic `source' at the origin, and apply this boundary condition in Section \ref{sec:continuity} to analyze the continuity equation governing probability conservation, and in Section 
 \ref{sec:wf} to the `bulk' wavefunction in both the sub- and supercritical regimes.  Since the wavefunction is singular at the source, we implement renormalization of the effective source-bulk coupling in Section \ref{sec:RG} that leads to the emergence of a scale anomaly. We also calculate the reflection and transmission probabilities; as these are observable physical quantities they should remain invariant under the RG flow. Our expressions for the reflection and transmission probabilities are therefore expressed in terms of RG invariant parameters. We finish with some perspectives and conclusions in Section \ref{sec:conclusion}.  Note that in the rest of this paper we specialize to the wavefunction in the scattering regime ($E>0$). However, this leads to the same basic critical behaviour and RG flow as the bound state case ($E<0$).

\section{Hamiltonian for the inverse square problem}
\label{sec:inversesqH}

The Hamiltonian we will work with is given by
\begin{equation} 
H = \frac{P^{2}}{2m} - \frac{g}{Q^{2}} .
\label{eq:invsqH}
\end{equation}
As noted above, both the potential and the kinetic energy scale in the same way, and this holds true in any dimension, unlike, say, $\delta$-function interactions which only scale in this way in two dimensions (in this paper we consider the 1D case). 
Classically, the conserved quantity corresponding to the continuous scale invariance of the inverse square Hamiltonian is the generator of the scale transformations $D = QP$, 
\begin{equation}
 \frac{dD}{dt} = 2H \label{almost_conservation_law} 
 \end{equation} 
 which is conserved if energy is zero, and is sometimes termed an \textit{almost} conservation law \cite{rajeev2013advanced}. In terms of the dimensionless parameter $\alpha= 2m g/\hbar^2$,  fall to the centre occurs when $\alpha > \alpha_{c} = 1/4 $ (supercritical regime). In this regime neither the boundedness of the Hamiltonian nor normalizability turn out to be good criteria for selecting either one of the two eigenfunctions over the other, as demonstrated by Burgess \textit{et al.} in \cite{cliffppeft1}.
 Furthermore,  the energy in the supercritical regime is unbounded from below. This is demonstrated explicitly in Ref.\ \cite{gupta1993renormalization}  by using a cleverly chosen  trial wavefunction to show  that 
\begin{equation}
 \alpha  \int \frac{\lvert \chi(Q)\rvert^{2}}{Q^{2}} dQ >\int \lvert \chi'(Q)\rvert^{2} dQ
 \end{equation}
 when $\alpha > \alpha_{c}$.  In 3D, the classical dynamics in the supercritical case gives rise to an unstable trajectory which spirals to the centre. 
 

 According to von Neumann's theorem, when a Hamiltonian is essentially self-adjoint the eigenvalue problem is mathematically well-posed. When the Hamiltonian is not self-adjoint one can attempt to prune the unsavoury parts by applying a self-adjoint extension.  It turns out that the inverse square problem is only self-adjoint for the strongly repulsive case and is not self-adjoint for any attractive potential, i.e. when $\alpha >0$.  While self-adjoint extensions can be applied in both the sub- and supercritical cases (like in the Efimov problem which is in the supercritical regime), they are not unique and so there is some arbitrariness involved \cite{capri1977self}. Furthermore, they do not necessarily remove the problem of unboundedness from below  \cite{gupta1993renormalization,basu2001,phdthesisGopalakrishnan}.  
  In the next section we introduce the PPEFT method which uses a source particle at the origin as a transparent way of choosing an appropriate boundary condition  \cite{cliffppeft1}. The source could be unitary (self-adjoint extension) or nonunitary (complex extension), either of which could be physically acceptable, depending on the situation. When combined with renormalization the results are independent of the regulator.  PPEFT gives the same results as other methods (in particular, it also does not remove unboundedness from below), but has the advantage of being intuitive and systematic.
  
\section{Point particle effective field theory and boundary condition}
\label{sec:ppeft}
 PPEFT starts from an action and this allows us to specify the physics and symmetries of the problem rather than imposing an arbitrary cut-off.  The total action is written as $S_{\mathrm{total}} = S_{B} + S_{p}$, where $S_{B}$ is the action for the Schr\"{o}dinger bulk field $\chi(Q)$ describing the long distance, low energy physics, and is given by 
\eq S_{B} = \int dt~dQ \left[\frac{\mbf{i}\hbar}{2}\left(\chi^{*}\partial_{t}\chi - \chi\partial_{t}\chi^{*}\right) -\lp\frac{\hbar^{2}}{2m}\lvert\nabla \chi \rvert^{2} + V(Q)\lvert\chi\rvert^{2} \rp \right] \qe
with $V(Q) = -\frac{g}{Q^{2}} $, and $S_{p}$ is the action describing the coupling between the bulk field  and the microscopic (short distance, high energy) source localized around $Q = 0$
\begin{equation} 
  S_{p} = \int dt~dQ \mathcal{L}_{p}(\chi^{*},\chi)\delta(Q).
\end{equation}  
The key to PPEFT is a series expansion of the lagrangian density  $\mathcal{L}_{p}(\chi^{*},\chi) = -h \chi^{*}\chi + \cdots $ where higher terms contain higher powers of $\chi$ and $\chi^{*}$ and/or their derivatives. This can be viewed as analogous to a multipole expansion where successive terms build in more information about the source but are less important at large distances. Thus, the coupling $h$ can be considered to be the `monopole' moment, and since each term must have the same total dimension, the multipole moments of higher terms will have correspondingly higher dimensions. Referring to Figure \ref{fig2}, the relevant scale for the source particle is $r$ and therefore the higher order moments will generically be proportional to $r/a$ raised to some power and hence smaller, allowing us to build in finer details about source-bulk coupling in a controlled fashion. In this paper we will not use the full power of PPEFT and will only retain the leading term meaning that we have just one coupling parameter $h$.

 The field equations can be obtained by extremizing the action $\delta
 S = 0$, to obtain the time dependent Schr\"{o}dinger equation 
 \begin{equation}
\lp - \frac{\partial^{2}}{\partial Q^{2}} + U(Q)\rp\chi = \mbf{i} \frac{2m}{\hbar} \frac{\partial \chi}{\partial t} \label{Full_inversesq_Eq}
 \end{equation}
 with 
 \begin{equation}
 U(Q) = -\frac{\alpha}{Q^{2}} + \lambda\delta(Q),
 \end{equation}
  where $\lambda=2m h /\hbar^2$ is the redefined source-bulk coupling constant. We see that the point particle effective action has modified the potential $V(Q)$ by adding a Dirac $\delta$-function, as previously claimed in Eq.\ (\ref{eq:schrod_with_delta}). 
  
The boundary condition is obtained by integrating the Schr\"{o}dinger equation over the infinitesimal region $-\epsilon\leq Q\leq \epsilon $, which gives 
\begin{equation}
\lambda = \left[\frac{\partial \ln(\chi)}{\partial Q}\right]^{Q = \epsilon}_{Q = -\epsilon} \label{boundary_condition_eps} .
\end{equation} 
The length scale $\epsilon$ is a cut-off or regulator which should be much shorter than the observable length scale $a$, but much larger than the size of the source $r$ located at the origin because the PPEFT action does not converge at distances where $\epsilon \approx r$. In other words, we need the hierarchy $a \gg \epsilon \gg r$ to be obeyed [see  Fig.~\ref{fig2}]. Furthermore, physical observables like reflection and transmission probabilities should both be finite and independent of the regularization scale $\epsilon$, and this is where renormalization of the source-bulk coupling $\lambda$ comes in. However, before coming to the renormalization there are two other tasks we can accomplish using the boundary condition given in (\ref{boundary_condition_eps}). The first is to use it in combination with the continuity equation to analyze probability conservation and the second is to apply it to the wavefunction.

\section{Continuity equation}
\label{sec:continuity}

The continuity equation is a basic tool for analyzing probability conservation. In a hermitian system we expect probability to be conserved, but in a nonhermitian system such basic properties can be violated.  In this section we show how probability conservation is determined  by the source-bulk coupling constant $\lambda$ by evaluating the probability current at the boundaries $Q=\pm \epsilon$. 

Proceeding from the Schr\"{o}dinger equation and its complex conjugate  in the usual way we find the continuity equation
\begin{equation}
 \partial_{t}\rho + \nabla.J = \frac{2}{\hbar}\rho\Im \left[ -\frac{g}{Q^{2}} + \frac{\hbar^2 \lambda}{2m} \delta(Q) \right] 
 \label{eq:continuity}
 \end{equation}
where $\rho$ is the probability density, and $J$ is the standard probability current
\begin{equation}
 J = \frac{\mbf{i}\hbar}{2m}\lp\chi\partial_{Q} \chi^{*} - \chi^{*}\partial_{Q}\chi\rp  .
 \end{equation}
The term in the square brackets in Eq.\ (\ref{eq:continuity}) is the total potential term in the Schr\"{o}dinger equation. Since $g$ is assumed to be real, the first term will make no contribution, but the second term will if $\lambda$ is complex.

The boundary condition Eq.~(\ref{boundary_condition_eps}) can be used to calculate the net probability current out of the origin
\begin{equation} 
J(\epsilon) - J(-\epsilon) = \frac{\mbf{i}\hbar}{2m} (\lambda^{*} - \lambda)\chi^{*}(\epsilon)\chi(\epsilon) .
\end{equation} 
We can therefore refine our condition for probability conservation violation to saying that $\lambda$ must be complex (nonhermitian Hamiltonian) and the probability for finding the particle at the source must be nonzero. The origin will be a sink if $\Im(\lambda) < 0 $ or a source if $\Im(\lambda) > 0$, as pointed out in \cite{plestid2018fall}.

\section{Wavefunction of the inverse square Hamiltonian in 1D and the boundary condition}
\label{sec:wf}

The time-independent Schr\"{o}dinger equation with energy eigenvalue $E=\hbar^{2} k^2 /2m > 0$ and purely inverse square potential is given by
\begin{equation}
\lp -\frac{d^{2}}{dQ^{2}} - \frac{\alpha}{Q^{2}}\rp\chi(Q) = k^2 \chi(Q) . \label{inverse_sq_schroeqn}
\end{equation} 
Putting $z = kQ$ and $\chi(Q) = \sqrt{z}u(z)$,  this is transformed to the Hankel differential equation
 \begin{equation} 
 z^{2}u'' + zu' + (z^{2} - \sigma^{2})u = 0 \label{Hankel_diff_eqn1} 
 \end{equation}
where $\sigma^{2} = 1/4 - \alpha$. The parameter $\sigma$ is real in the subcritical regime, but in the supercritical regime it is an imaginary number which we write as $\sigma = \pm\mbf{i}\zeta$,  where $\zeta \in \mathbb{R}$. It is notable that, unlike $\sigma^2$, the eigenvalue $k^2$ does not appear as an explicit parameter in Eq.\ (\ref{Hankel_diff_eqn1}) but instead occurs only as a scaling factor in the argument of the solutions.  We also note in passing that the Hankel differential equation is invariant under $z \rightarrow -z$,  i.e. under parity $\mathcal{P}$, and also under $\sigma \rightarrow -\sigma$.

The independent solutions to the Hankel equation are the two Hankel functions $H_{\sigma}^{(1)}(z)$ and $H_{\sigma}^{(2)}(z)$ whose properties, including asymptotics, are summarized in the Appendix. From these it can be inferred that $H_{\sigma}^{(1)}(z)$ asymptotes at large $z$ to a right moving wave and $H_{\sigma}^{(2)}(z)$ asymptotes to a left moving wave. Translating back to the original wavefunction $\chi(Q)$, and assuming an initial wave coming in from the right, the general solution to the scattering problem can be written as
\begin{eqnarray}
\chi_{\mbf{in+ref}}(Q) &=& \sqrt{kQ}\lp H_{\sigma}^{(2)}(kQ) + R H_{\sigma}^{(1)}(kQ)\rp,Q\geq\epsilon \\
\chi_{\mbf{trans}}(Q) &=& T\sqrt{kQ}H_{\sigma}^{(2)}(kQ) ~,~ Q\leq -\epsilon
\end{eqnarray} 
where, $R$ and $T$ are the reflection and transmission amplitudes to be determined by the boundary condition at $Q = \pm \epsilon$.

\subsection{Boundary condition for the subcritical case}

 In the subcritical case $\sigma \in \mathbb{R}$.  We can find a relation between R and T by demanding continuity of the wavefunction near the origin $\chi_{\mbf{in+ref}}(\epsilon) = \chi_{\mbf{trans}}(-\epsilon)$, yielding
\begin{equation}
R + \mbf{i}T \exp(\mbf{i}\pi \sigma)  = -\frac{H_{\sigma}^{(2)}(k\epsilon)}{H_{\sigma}^{(1)}(k\epsilon)}  
\end{equation}
where we have used the reflection identities of the Hankel functions given in the Appendix.
 For $k\epsilon \ll 1$, this relation becomes
\begin{equation}
 R + \mbf{i}T \exp(\mbf{i}\pi \sigma)  \approx \frac{1-X\exp(\mbf{i}\pi \sigma)}{1-X\exp(-\mbf{i}\pi \sigma)}\label{condition_RandTsub}
 \end{equation}
where 
\begin{equation}
 X = \frac{\Gamma(1-\sigma)}{\Gamma(1+\sigma)}\lp\frac{k\epsilon}{2}\rp^{2\sigma} .  
\end{equation} 
 We now consider the boundary condition given in Eq.\ (\ref{boundary_condition_eps}). When expressed in terms of the scattering solution it reads
 \begin{equation}
 \lambda = \frac{\partial \ln[\chi_{\mbf{in+ref}}(\epsilon)]}{\partial Q} - \frac{\partial \ln[\chi_{\mbf{trans}}(-\epsilon)]}{\partial Q} 
 \end{equation}
and we find the following expression for the coupling constant 
\begin{equation}
\lambda = \frac{1}{\epsilon}\lp 1 - \sigma\left[\frac{1+X\exp(\mbf{i}\pi \sigma)-R(1+X\exp(-\mbf{i}\pi \sigma))}{1-X\exp(\mbf{i}\pi \sigma)-R(1-X\exp(-\mbf{i}\pi \sigma))}+\frac{1+X\exp(-\mbf{i}\pi \sigma)}{1-X\exp(-\mbf{i}\pi \sigma)}\right]\label{lambda_func_eps_first_sub} \rp .
 \end{equation} 
 
In general $\lambda$ is complex and hence breaks $\mathcal{PT}$ symmetry. However, in this paper we focus on the fixed points of the RG flow and we will show in Section \ref{sec:RG} that in the subcritical case the UV and IR fixed points are real and hence preserve $\mathcal{PT}$ symmetry. Expanding Eq.~(\ref{lambda_func_eps_first_sub}) in powers of $X$ in the small $k\epsilon$ regime we obtain 
\begin{eqnarray}
  \Lambda \approx -2\sigma\left[1 + X\exp(\mbf{i}\pi \sigma)\lp\frac{1 - R\exp(-2\mbf{i}\pi \sigma)}{1-R}\rp+
   X\exp(-\mbf{i}\pi \sigma) + \mathcal{O}(X^{2})\right]  \label{coupling_small_eps}
   \end{eqnarray}
  where in order to simplify the expression we have defined 
\begin{equation}
\Lambda \equiv  2(\lambda\epsilon - 1)  \ .
\end{equation}

\subsection{Boundary condition for the supercritical case}

A very similar calculation to that given above can be performed to find the coupling constant $\lambda$ in the supercritical case, the only difference being that $\sigma$ is purely imaginary. Of the two possibles choices for the square root, we choose $\sigma = - \mathrm{i} \zeta$.  We find
\begin{eqnarray}
\chi_{\mbf{in+ref}}(Q) =& \frac{\sqrt{kQ} \lp\frac{-\lp\frac{kQ}{2}\rp^{\mbf{i}\zeta}}{\Gamma(1+\mbf{i}\zeta)} +\frac{\exp(\pi \zeta)\lp\frac{kQ}{2}\rp^{-\mbf{i}\zeta}}{\Gamma(1-\mbf{i}\zeta)}\rp + R \lp\frac{\lp\frac{kQ}{2}\rp^{\mbf{i}\zeta}}{\Gamma(1+\mbf{i}\zeta)} -\frac{\exp(-\pi \zeta)\lp\frac{kQ}{2}\rp^{-\mbf{i}\zeta}}{\Gamma(1 - \mbf{i}\zeta)}\rp}{\sinh(\pi \zeta)},\,Q\geq \epsilon \label{psi+} \\
\chi_{\mbf{trans}}(Q) =& \frac{T\sqrt{kQ}}{\sinh(\pi \zeta)}\lp\frac{-1}{\Gamma(1+\mbf{i}\zeta)}\lp\frac{kQ}{2}\rp^{\mbf{i}\zeta} +\frac{\exp(\pi \zeta)}{\Gamma(1-\mbf{i}\zeta)}\lp\frac{kQ}{2}\rp^{-\mbf{i}\zeta}\rp ~,~ Q\leq -\epsilon \ . \label{psi-}
\end{eqnarray}
Demanding continuity near the origin yields the relation
\begin{equation}
 R + \mbf{i}T \exp(\pi \zeta)  = -\frac{H_{-\mbf{i}\zeta}^{(2)}(k\epsilon)}{H_{-\mbf{i}\zeta}^{(1)}(k\epsilon)}  \ .
\end{equation}  
 For small $k\epsilon \ll 1$, this relation can be written 
 \begin{equation}
  R + \mbf{i}T \exp(\pi \zeta)  \approx \frac{1-X\exp(\pi \zeta)}{1-X\exp(-\pi \zeta)}\label{condition_RandT}  \ .
\end{equation}   
The coupling constant in the super critical case is then given by : 
  \begin{equation}
  \lambda = \frac{1}{\epsilon}\lp 1+ \mbf{i}\zeta\left[\frac{1+X\exp(\pi \zeta)-R(1+X\exp(-\pi \zeta))}{1-X\exp(\pi \zeta)-R(1-X\exp(-\pi \zeta))}+\frac{1+X\exp(-\pi \zeta)}{1-X\exp(-\pi \zeta)}\right]\label{lambda_func_eps_first_super} \rp
  \end{equation}
where 
\begin{equation}
X = \frac{\Gamma(1-\sigma)}{\Gamma(1+\sigma)}\lp\frac{k\epsilon}{2}\rp^{2\sigma}  \ .
\end{equation} 
  Expanding Eq.~(\ref{lambda_func_eps_first_super}) in powers of $X$ in the small $k\epsilon$ regime we obtain
  \begin{eqnarray}
  \Lambda \approx 2\mbf{i}\zeta\left[1 + X\exp(\pi \zeta)\lp\frac{1 - R\exp(-2\pi \zeta)}{1-R}\rp+
   X\exp(-\pi \zeta) + \mathcal{O}(X^{2})\right] \ . \label{coupling_small_eps}
   \end{eqnarray}
   We now proceed to renormalize the source-bulk coupling constant so that the reflection and transmission probabilities do not depend upon the regulator scale $\epsilon$ and are finite. We also discuss the fixed point merger of the renormalization group as a $\mathcal{PT}$ symmetry breaking transition as the strength of the inverse square potential $\alpha$ is tuned from the subcritical to supercritical regime.    
   
\section{Renormalization group}
\label{sec:RG} 

\subsection{Renormalization group and fixed point merger exhibiting a $\mathcal{PT}$ transition}

If we blindly let $\epsilon \rightarrow 0$ we run into trouble because the wavefunction develops singularities at small $Q$ [see Eq.\ (\ref{eq:shortrange_chi})]. PPEFT then breaks down  because the bulk wavefunction is singular at the source located at $Q = 0$. To deal with this  we must renormalize the source-bulk coupling constant $\lambda$. Following Ref.\ \cite{cliffppeft1}, we start by writing down the RG flow equation for the coupling constant $\lambda$, or more conveniently its close relative $\Lambda =  2(\lambda\epsilon - 1) $. This can be obtained by taking the derivative with respect to $\epsilon$ of the above equation for $\Lambda$, keeping all observables  (such as $k$) fixed. After some algebra this gives
\begin{equation}
\epsilon\frac{d}{d\epsilon}\lp\frac{\Lambda}{2\sigma}\rp = \sigma\lp 1-\lp\frac{\Lambda}{2\sigma}\rp^{2}\rp . \label{RG_evolution}
\end{equation}
This equation determines how the coupling constant $\lambda$ must depend on the regulator $\epsilon$ to renormalize any divergences such that physical quantities are independent of $\epsilon$. The non-trivial zeroes of the right hand side at $\Lambda=\pm 2 \sigma$ correspond to the fixed points of the flow. At fixed points the theory is scale invariant. An important consequence of the RG running of the coupling of the Dirac $\delta$-function is as follows: the vanishing of the coefficient of the $\delta$-function ($\lambda = 0$) only happens at $\Lambda = -2$, which is not a fixed point unless $\sigma = \pm 1 $, but this value of $\sigma$ is impossible to realize with an attractive inverse square potential because $\sigma=\sqrt{1/4 - \alpha}$. Thus, a $\delta$-function term is inevitable: the flow will always produce one.


The above RG evolution equation, considered as a first order differential equation, has a relatively simply quadratic right hand side and can be integrated analytically. In terms of the initial condition $\lambda(\epsilon_{0}) \equiv \lambda_{0}$, one finds
 \begin{equation}
\frac{\Lambda}{2\sigma} = \frac{\frac{\lambda_{0}}{2\sigma}+\tanh(\sigma\ln(\epsilon/\epsilon_{0}))}{1+\frac{\lambda_{0}}{2\sigma}\tanh(\sigma\ln(\epsilon/\epsilon_{0}))}\label{rg_flow} .
\end{equation}

\subsubsection{Supercritical case} 
 
\begin{figure*}
\begin{center}
\subfloat[]{\includegraphics[width = 0.5\columnwidth ]{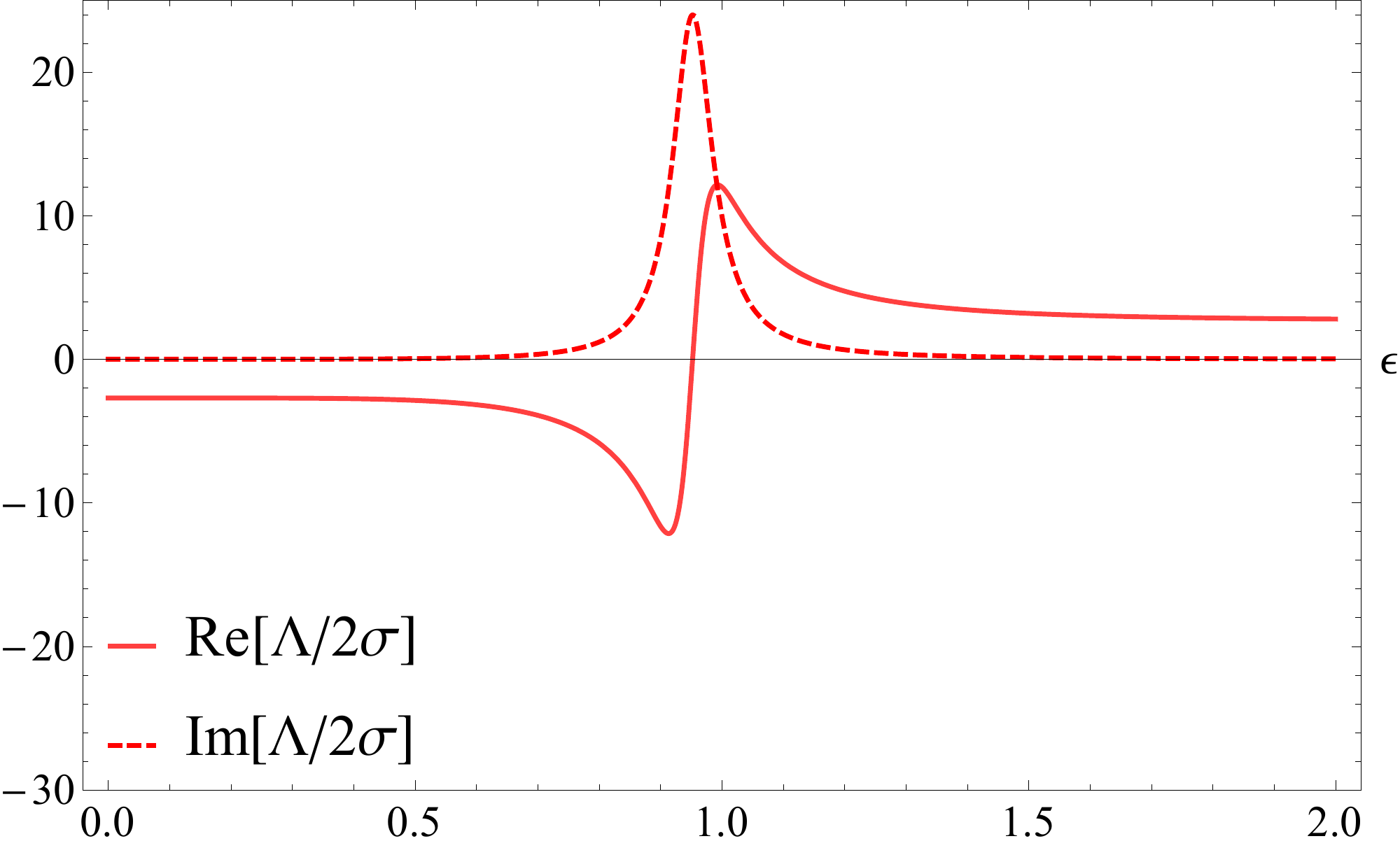}}
\subfloat[]{\includegraphics[width = 0.5\columnwidth]{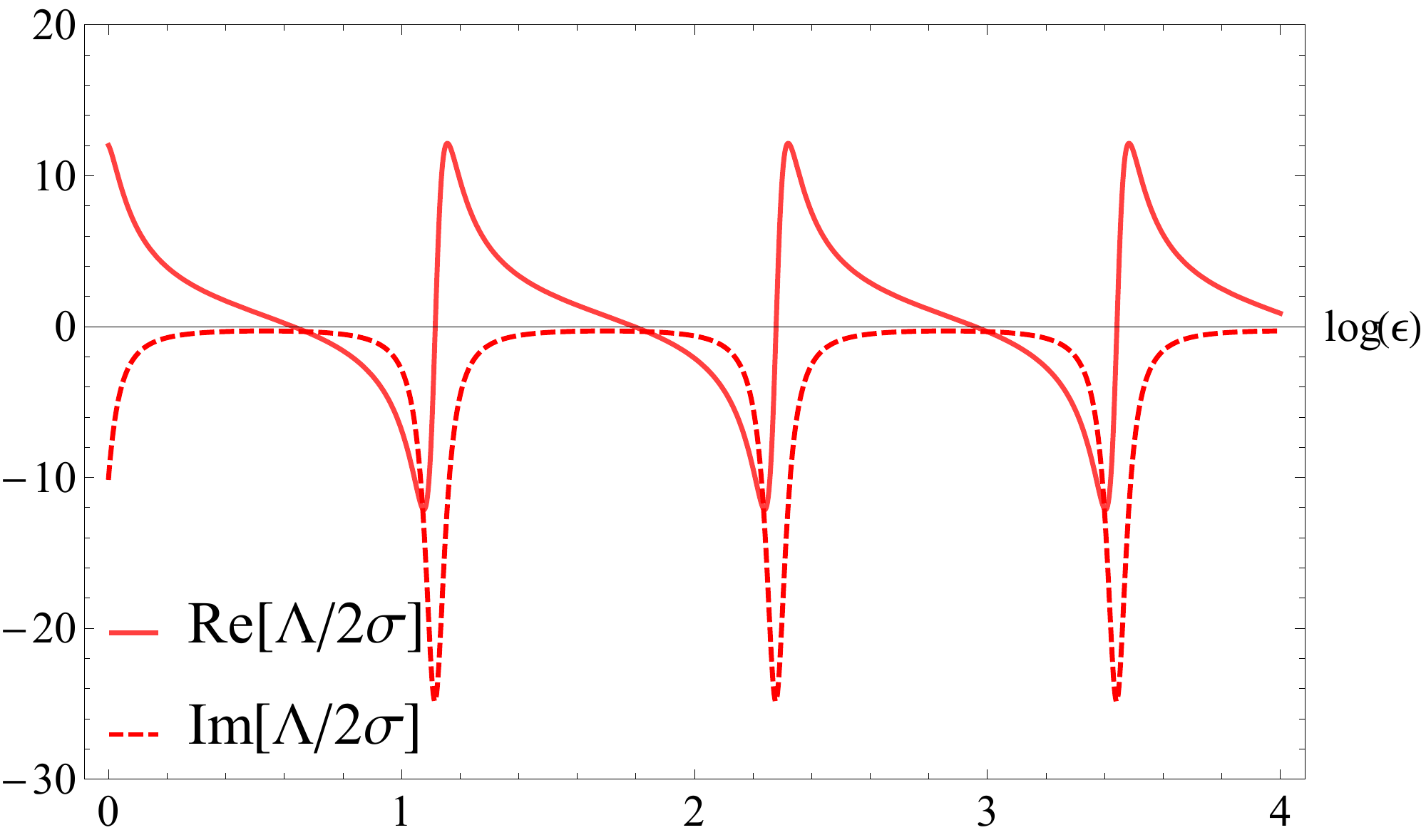}}
\end{center}
 \caption{Panel (a) shows the real and imaginary parts of the RG flow as a function of $\epsilon$ for the subcritical case which means that $\sigma$ is real. Panel (b) shows the real and imaginary parts of the RG flow in the supercritical case which means that $\sigma$ is imaginary. In the supercritical case the flow exhibits a log periodic behaviour.}
 \label{fig3}
 \end{figure*}

\begin{figure*}
\begin{center}
\subfloat[]{\includegraphics[width = 0.5\columnwidth ]{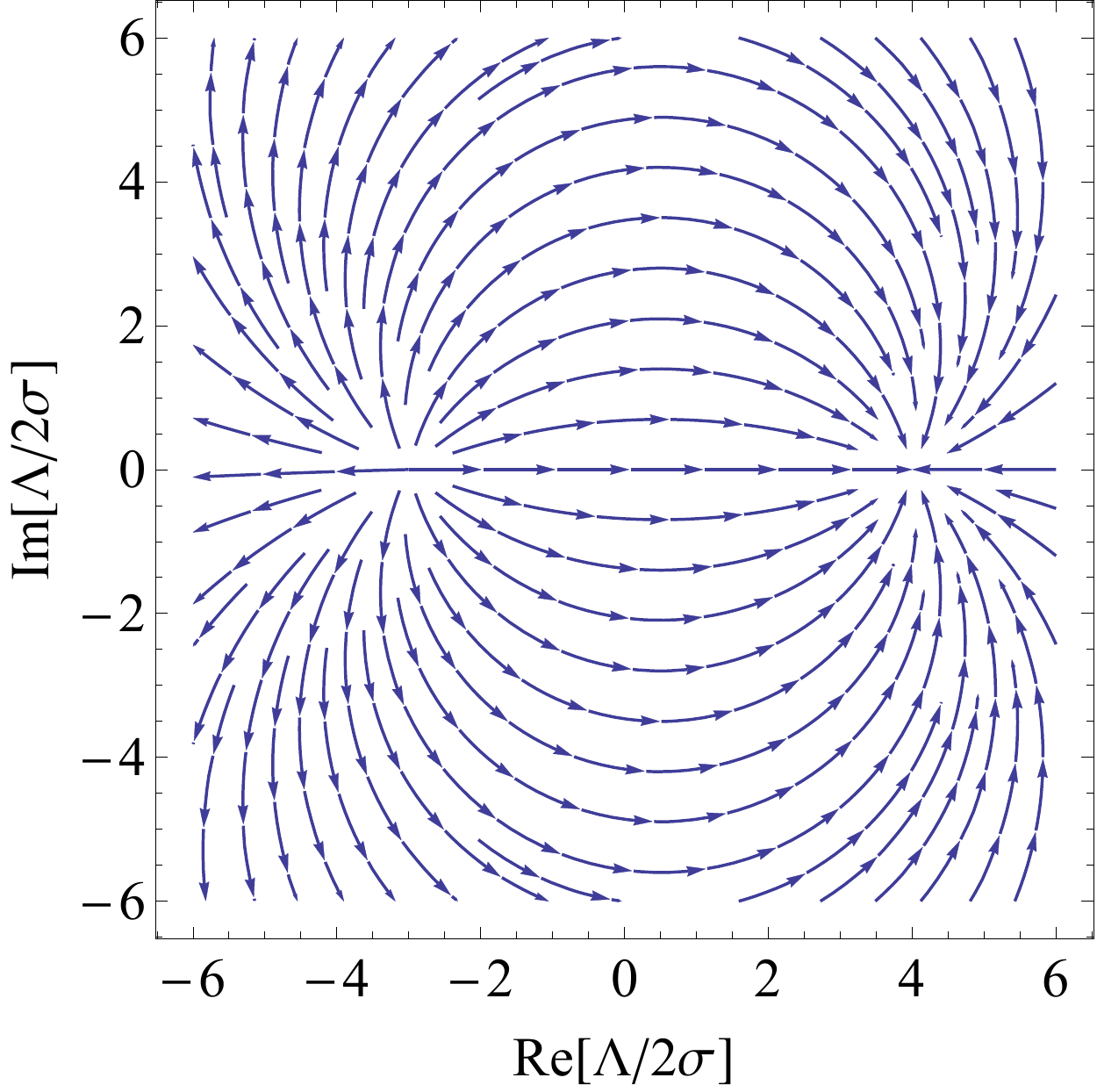}}
\subfloat[]{\includegraphics[width = 0.5\columnwidth]{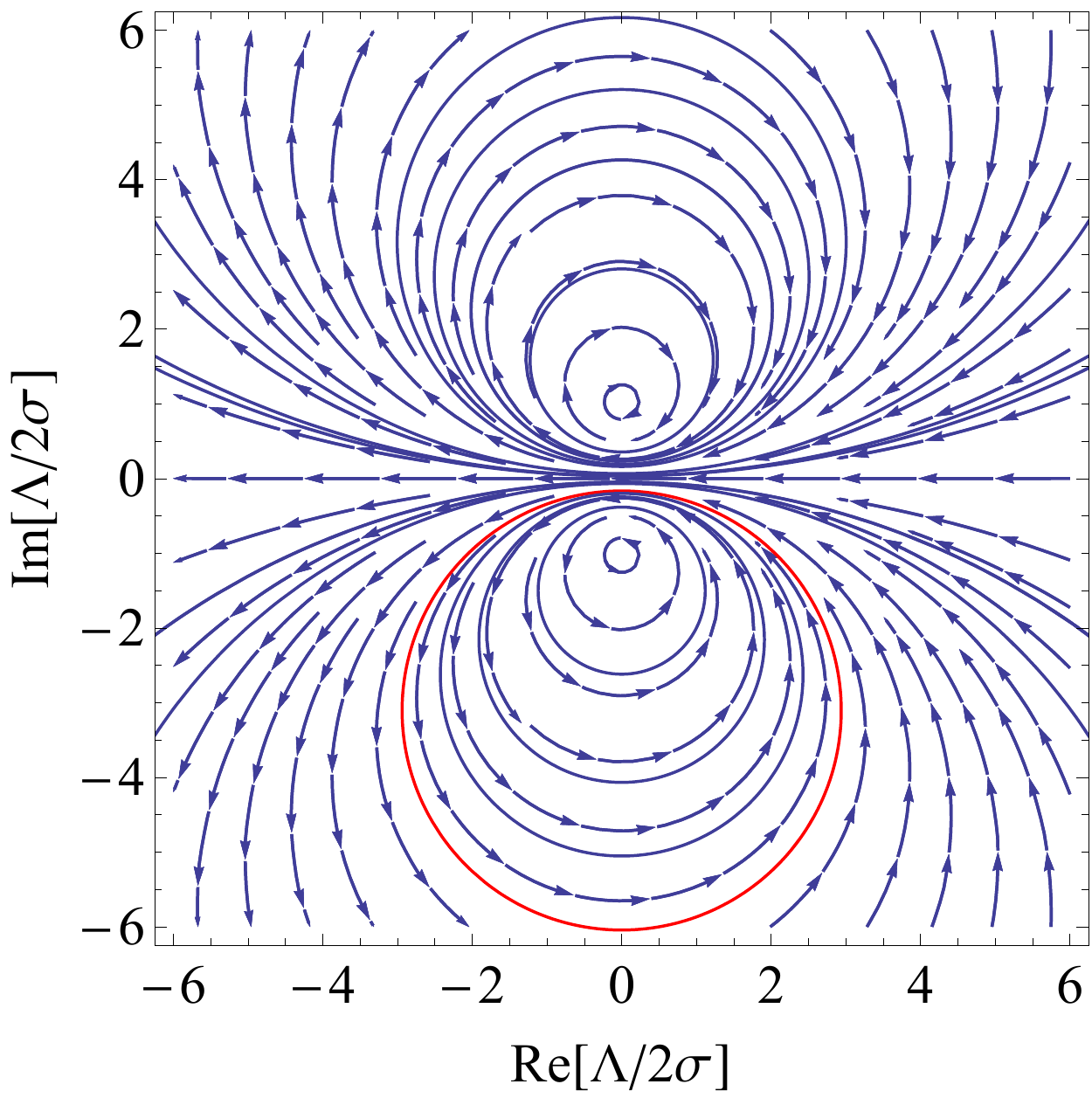}}
\end{center}
 \caption{Panel (a) shows the phase portrait of the RG flow in the subcritical case (real $\sigma$) which has two real fixed points, one stable and the other unstable. Panel (b) shows a limit cycle behaviour in the supercritical case (imaginary $\sigma$). Each RG trajectory picks a length scale $\epsilon_{*}$ when $\Re[\hat{\lambda}] = 0$. Arrows indicate the direction of flow. Each trajectory is characterized by the values of $\epsilon_{*}$ and $y_{*}$, where $y_{*} = \Im[\lambda(\epsilon_{*})]$.}
 \label{fig4}
 \end{figure*}
 In the supercritical case when $\alpha > 1/4$, the fixed points for Eq.\ (\ref{RG_evolution}) are given by $\Lambda = \pm 2\sigma = \pm 2\mbf{i}\zeta$, $\zeta \in \mathbb{R}$. Thus, there are two fixed points and they are purely imaginary complex conjugates of each other.  The RG evolution when $\sigma$ is imaginary is dealt with in detail in \cite{plestid2018fall}, but the resulting flow of the coupling $\Lambda$ as $\epsilon$ is increased is shown in Fig. \ref{fig3}(b). The limit cycle behaviour of the trajectories is illustrated in Fig. \ref{fig4}(b),  and combined with the logarithmic derivative in Eq.\ (\ref{RG_evolution}) means that the  flow exhibits a log periodic behaviour as a function of $\epsilon$. Note that each trajectory can be uniquely labelled by, for example, its value where it crosses the imaginary axis ($\Re [ \Lambda] = 0$) and this label is then an RG invariant. In fact, the flow picks a scale $\epsilon_{*}$ at this point  and hence breaks continuous scale invariance, exhibiting discrete scale invariance instead.  The limit cycle behaviour means that trajectories that start away from the fixed points can never reach them. Physically, these fixed points correspond to the (scale invariant) scenarios of a perfect sink when $\Im(\lambda) < 0$ and a perfect source when $\Im(\lambda) > 0$ and because they are complex they break $\mathcal{PT}$ symmetry.

To calculate the reflection coefficient we can use the small $\epsilon$ expansion of Eq.~(\ref{rg_flow}) 
\begin{equation}
\frac{\Lambda}{2\sigma} \approx -1-2\lp\frac{\epsilon}{\epsilon_{*}}\rp^{2\sigma}, \quad ~(\epsilon \ll \epsilon_{*})
\end{equation} 
and substitute it into Eq.~(\ref{coupling_small_eps}) to obtain 
\begin{equation}
 R = \frac{X_{*}\cosh(\pi \zeta)-1}{X_{*}\exp(-\pi \zeta)-1} \qe where, \eq X_{*} = \frac{\Gamma(1-\sigma)}{\Gamma(1+\sigma)}\lp\frac{k\epsilon_{*}}{2}\rp^{2\sigma} 
\end{equation} 
 and using the small $k\epsilon$ limit in Eq.~(\ref{condition_RandT}) finally yields
 \begin{equation}
 T = -\mbf{i}\exp(-\pi \zeta)(1-R) = \frac{\mbf{i}X_{*}\exp(-\pi \zeta)\sinh(\pi \zeta)}{X_{*}\exp(-\pi \zeta)-1} . \label{non_unitarity}
 \end{equation} 
 Note that all observables are expressed entirely in terms of RG invariant quantities. 

\subsubsection{Subcritical case}
 In the subcritical case $\sigma \in \mathbb{R}$. The solution of the RG flow equation as a function of $\epsilon$ for this case  is shown in Fig. \ref{fig3}(a). If we chose to start with real $\Lambda$ as an initial condition, the coupling will stay real as the flow evolves: hermiticity is itself an RG invariant. The RG phase portraits in sub- and supercritical cases are strikingly different: every RG trajectory in the phase portrait for the subcritical case flows from the UV fixed point $\Lambda = -2\sigma$ to the IR fixed point $\Lambda = +2\sigma$ as shown in Fig. \ref{fig4}(a). The two real fixed points correspond to scale invariant phases and are also  $\mathcal{PT}$ symmetric.  If we tune the strength of the inverse square potential $\alpha$ the critical points merge at $\alpha_{c} = 1/4$. Hence, we view the fixed point merger as a form of $\mathcal{PT}$ symmetry breaking transition. 
   
 To calculate the reflection coefficient, we follow a similar procedure to the supercritical case. We use the  small $\epsilon$ expansion of Eq.~(\ref{rg_flow})
\begin{equation}
\frac{\Lambda}{2\sigma} \approx -1-2\lp\frac{\epsilon}{\epsilon_{*}}\rp^{2\sigma}, \quad ~(\epsilon << \epsilon_{*})
\end{equation}  
to obtain \cite{cliffppeft1}
\begin{equation}
 R = \frac{X_{*}\cos(\pi \sigma)-1}{X_{*}\exp(-i\pi \sigma)-1} \qe where \eq X_{*} = \frac{\Gamma(1-\sigma)}{\Gamma(1+\sigma)}\lp\frac{k\epsilon_{*}}{2}\rp^{2\sigma}.
\end{equation}  
Finally, the transmission coefficient can be expressed as
\begin{equation}
T = -\mbf{i}\exp(-i\pi \sigma)(1-R) = \frac{\mbf{i}X_{*}\exp(-i\pi \sigma)\sin(\pi \sigma)}{X_{*}\exp(-i\pi \sigma)-1} .
\end{equation}

\section{Conclusion}
\label{sec:conclusion}

In this work we have argued that fall to the centre is a form of $\mathcal{PT}$ symmetry breaking transition if one focuses on the fixed point structure of the RG flow. The seemingly simple inverse square Hamiltonian presents some subtle difficulties and by itself is not a fully defined problem: one must impose a boundary condition representing additional microscopic `source' physics at the origin to make the eigenvalue problem well posed. We use the PPEFT tools of ref.~\cite{cliffppeft1} to derive the boundary condition for the 1D case in both the sub- and supercritical regimes. This amounts to adding an inevitable Dirac $\delta$-function at the origin. Depending upon the nature of the physics at the origin, one can choose a unitary or a non-unitary boundary condition and the Hamiltonian becomes non-hermitian when a non-unitary boundary condition is implemented. The source-bulk coupling $\lambda$ is evaluated by a boundary condition on the wavefunction at the length scale  $\epsilon$ (which is a regulator scale that is arbitrary). The RG flow of the source-bulk coupling shows several interesting properties, and in this paper we show that the fixed point merger that occurs as the strength of the inverse square potential $\alpha$ is tuned from subcritical to supercritical  \cite{moroz2010nonrelativistic} is also a $\mathcal{PT}$ symmetry breaking transition. In particular, the two real RG fixed points of the source-bulk coupling $\lambda$ in the subcritical case ($\alpha < 1/4$), which are attractive/repulsive and preserve $\mathcal{PT}$ symmetry, merge as we tune $\alpha$ to the critical value $\alpha_{c} = 1/4$. As the strength of the potential is further increased above the critical value, i.e.\ when $\alpha > 1/4$, we enter the supercritical regime and the fixed points disappear into the complex plane, breaking $\mathcal{PT}$ symmetry. Thus, while the system can be both scale invariant (at a fixed point) {\it and} unitary for $\alpha < 1/4$, one must choose {\it either} scale invariance {\it or} unitarity for $\alpha > 1/4$. 
We do not claim that fall to the centre is an exact realization of the standard $\mathcal{PT}$ breaking transition \cite{bender1998real, bender1999pt, bender2002complex}, but it has many similar elements. One of the differences is that, although the Hamiltonian is generically nonhermitian in both phases, it can also be chosen to be hermitian ($\lambda$ real) in both phases, rare though these trajectories are. It is, however, nonselfadjoint in both phases if one does not regulate it. The precise role of $\mathcal{P}$ is also elusive (although the Hamiltonian and the boundary condition both always appear to be $\mathcal{P}$ symmetric), and this remains a topic for further investigation.\\


We acknowledge the support of the Natural Sciences and Engineering Research Council of Canada
(NSERC) [Ref.\ No.\ RGPIN-2017-06605 and SAPIN-2020-00045]. Research at the Perimeter Institute is supported in part by the Government of Canada through NSERC and by the Province of Ontario through MRI.

\appendix

\section{Properties and asymptotics of Hankel functions}
\label{A3}
In this appendix we state some of the properties of the Hankel function that are used in the paper. The Hankel differential equation is given by: \eq z^{2}u'' + zu' + (z^{2} - \sigma^{2})u = 0\label{Hankel_diff_eqn} \ . \qe  Hankel functions of first and second kind are defined by: \eq H_{\sigma}^{(1)}(z) = J_{\sigma}(z) + \mbf{i}N_{\sigma}(z)  \qe and \eq H_{\sigma}^{(2)}(z) = J_{\sigma}(z) - \mbf{i}N_{\sigma}(z) \qe where $J_{\sigma}$ is the Bessel function and $N_{\sigma}$ is the Neumann function.

The asymptotic large $z$ behaviour of Hankel functions is given by:
 \begin{equation}
  H_{\sigma}^{(1)}(z) \sim \sqrt{\frac{2}{\pi z}}\exp\left[\mbf{i}\lp z -\frac{\pi\sigma}{2}-\frac{\pi}{4}\rp\right] 
\end{equation}  
\begin{equation}
  H_{\sigma}^{(2)}(z) \sim \sqrt{\frac{2}{\pi z}}\exp\left[-\mbf{i}\lp z -\frac{\pi\sigma}{2}-\frac{\pi}{4}\rp\right] \ .
\end{equation}  
and they have the following reflection properties:
 \begin{equation}
H_{\sigma}^{(1)}(\exp(\mbf{i}\pi) z) = -\exp(-\mbf{i}\pi\sigma)H_{\sigma}^{(2)}(z)
\end{equation} 
\begin{equation}
H_{\sigma}^{(2)}(\exp(-\mbf{i}\pi) z) = -\exp(\mbf{i}\pi\sigma)H_{\sigma}^{(1)}(z) \ .
\end{equation}  
For small $z$, Hankel functions reduce to monomials:
\begin{equation}
H_{\sigma}^{(1)}(z) \approxeq \frac{1}{\mbf{i}\sin(\pi\sigma)}\lp\frac{1}{\Gamma(1-\sigma)}\lp\frac{z}{2}\rp^{-\sigma} -\frac{\exp(-\mbf{i}\pi\sigma)}{\Gamma(1+\sigma)}\lp\frac{z}{2}\rp^{\sigma}\rp
\end{equation}  

\begin{equation}
 H_{\sigma}^{(2)}(z) \approxeq \frac{1}{\mbf{i}\sin(\pi\sigma)}\lp-\frac{1}{\Gamma(1-\sigma)}\lp\frac{z}{2}\rp^{-\sigma} +\frac{\exp(\mbf{i}\pi\sigma)}{\Gamma(1+\sigma)}\lp\frac{z}{2}\rp^{\sigma}\rp 
\end{equation}


\end{document}